\documentclass[12pt]{article}
\usepackage{graphics}
\usepackage{amssymb}

\newcommand{\R}{\mathbb{R}}

\newcommand{\D}{\mathrm{d}}

\newcommand{\CC}{\mathcal{C}}

\newcommand{\EE}{\mathcal{E}}
\newcommand{\OO}{\mathcal{O}}
\newcommand{\PP}{\mathcal{P}}
\newcommand{\RR}{\mathcal{R}}
\newcommand{\SSS}{\mathcal{S}}
\newcommand{\im}{\mathrm{Im\,}}
\newcommand{\QED}{\mbox{\rule[-.5pt]{4pt}{7.5pt}}}
\newtheorem{claim}{Claim}[section]
\newtheorem{theorem}[claim]{Theorem}
\newtheorem{conjecture}[claim]{Conjecture}
\newtheorem{proposition}[claim]{Proposition}
\newtheorem{remark}[claim]{Remark}
\newtheorem{example}[claim]{Example}

\begin{document}

\title{An isoperimetric problem for leaky loops \\ and related
mean-chord inequalities}
\author{Pavel Exner}
\date{}
\maketitle
\begin{center}
{\small \em Department of Theoretical Physics, Nuclear Physics
Institute, \\ Academy of Sciences, 25068 \v{R}e\v{z} near Prague,
Czechia, and \\ Doppler Institute, Czech Technical
University, B\v{r}ehov{\'a} 7, \\
11519 Prague, Czechia \\ \rm exner@ujf.cas.cz}
\end{center}

\vspace{8mm}

\begin{quote}
\noindent {\small We consider a class of Hamiltonians in
$L^2(\R^2)$ with attractive interaction supported by piecewise
$C^2$ smooth loops $\Gamma$ of a fixed length $L$, formally given
by $-\Delta-\alpha\delta(x-\Gamma)$ with $\alpha>0$. It is shown
that the ground state of this operator is locally maximized by a
circular $\Gamma$. We also conjecture that this property holds
globally and show that the problem is related to an interesting
family of geometric inequalities concerning mean values of chords
of $\Gamma$. }
\end{quote}


\section{Introduction}

There is a small number of topics which can be regarded as
trademark for mathematical physics. One of them without any doubt
concerns relations between geometric properties of constraints
and/or interaction and extremal values of a spectral quantity;
classical examples are Faber-Krahn inequality \cite{Fa, Kr} or the
PPW-conjecture proved by Ashbaugh and Benguria \cite{AB}.

A common feature of these and analogous problems is that the
extremum is reached by shapes having a rotational symmetry. At the
same time, the nature of the extremum may be different. While a
ball \emph{minimizes} the principal eigenvalue of the Dirichlet
Laplacian among regions of a fixed volume, for non-simply
connected regions like annular strips or layers considered in
\cite{EHL, HKK}, built over a curve (surface) of a fixed length
(area), the circular shape is on the contrary a \emph{maximizer}.
A natural topological way to understand this difference becomes
smeared, however, when the particle is not localized by boundary
conditions but by a potential, a regular or singular one.

In this paper we consider such a problem associated with a class
of operators in $L^2(\R^2)$ which are given formally by the
expression
 \begin{equation} \label{formal}
 H_{\alpha,\Gamma}= -\Delta-\alpha\delta(x-\Gamma)\,,
 \end{equation}
where $\alpha>0$ and $\Gamma$ is a $C^2$ loop in the plane (see
below for exact assumptions) having a fixed length $L>0$. A
motivation to study these operators comes from the theory of
\emph{leaky quantum graphs} -- see \cite{BT, EI} and related
papers, a bibliography can be found in \cite{AGHH2} -- aiming at a
more realistic model of quantum wire structures which would take
quantum tunneling into account.

Our aim is to show that the ground-state energy of
$H_{\alpha,\Gamma}$ is (sharply) maximized when $\Gamma$ is a
circle. We will be able to prove that this property holds
\emph{locally} conjecturing its global validity. There are several
reasons why one may expect this result to be valid. On one hand,
we know from \cite{EY} that in the limit of strong coupling,
$\alpha \to\infty$, the ground-state dependence on $\Gamma$ is
given in the leading order by the lowest eigenvalue of the
operator $-\frac{\D^2}{\D s^2}- \frac14 \gamma(s)^2$ on
$L^2([0,L])$ with periodic boundary conditions where $\gamma$ is
the curvature of $\Gamma$, and the latter is easily seen to be
globally sharply maximized when $\gamma$ is constant along
$\Gamma$. On the other hand, by \cite{EN} the operator
$H_{\alpha,\Gamma}$ can be approximated in the strong resolvent
sense by point interaction Hamiltonians with the point
interactions equidistantly spaced along $\Gamma$ and properly
chosen coupling constants, and from \cite{Ex} we know that the
ground state of such an operator is locally maximized by a regular
polygon.

Needless to say, neither of the above observations proves the
desired result. The first one is global, but it holds only
asymptotically and we do not know whether the error term will not
spoil the inequality. The second argument holds for any $\alpha>0$
suggesting the local validity, but the polygons approximating the
circle do not have exactly the same lengths.

Our main tools in this paper are the generalized Birman-Schwinger
principle in combination with the convexity of Green's function.
They allow us to reformulate the problem in a purely geometric
way, in terms of \emph{mean value of chords} of arc segments of
$\Gamma$. Since such geometric inequalities are of an independent
interest, we discuss them in Sec.~\ref{geom} separately in a
broader context, including the discrete version which arose in
connection with the polygon problem treated in \cite{Ex}. Before
doing that, we will formulate in the next section the problem and
state our main result, Theorem~\ref{mainthm}, and provide the
mentioned reformulation in Sec.~\ref{bs}. After the discussion of
the inequalities we will finish the proof of Theorem~\ref{mainthm}
and present some concluding remarks.


\setcounter{equation}{0}
\section{Formulation and the main result} \label{main}

We will assume throughout that $\Gamma:\: [0,L]\to \R^2$ is
\emph{a closed curve, $\Gamma(0)= \Gamma(L)$, parametrized by its
arc length, which is $C^1$-smooth, piecewise $C^2$, and has no
cusps}\footnote{There are, of course, no local cusps under the
$C^1$ assumption, but we have not excluded self-intersections, so
the last requirement means that the curve meets itself at such a
point at a nonzero angle. In fact our main result can be pushed
through under a slightly weaker regularity assumption, namely that
$\dot\Gamma$ is absolutely continuous.}. Unless stated otherwise,
we will mean by the curve $\Gamma$ for simplicity both the above
mentioned function and its image in the plane. Furthermore, we
introduce the equivalence relation: $\Gamma$ and $\Gamma'$ belong
to the same equivalence class if one can be obtained from the
other by a Euclidean transformation of the plane. Spectral
properties of the corresponding $H_{\alpha,\Gamma}$ and
$H_{\alpha,\Gamma'}$ are obviously the same, and we will usually
speak about a curve $\Gamma$ having in mind the corresponding
equivalence class. It is clear that the stated regularity
assumptions are satisfied, in particular, by the circle, say
$\CC:= \{\,((L/2\pi)\cos s, (L/2\pi)\sin s):\: s\in[0,L]\,\}$, and
its equivalence class.

First of all we have to give a rigorous meaning to the operator
(\ref{formal}). Following \cite{BEKS, BT} we can do that in two
ways. The more general one is to consider a positive Radon measure
$m$ on $\R^2$ and $\alpha>0$ such that
\begin{equation} \label{basiccond}
(1+\alpha) \int_{\R^2} |\psi(x)|^2\, \D m(x) \le a \int_{\R^2}
|\nabla\psi(x)|^2\, \D x + b \int_{\R^2} |\psi(x)|^2\, \D x
\end{equation}
holds for all $\psi$ from the Schwartz space $\SSS(\R^2)$ and some
$a<1$ and $b$. The map $I_m$ defined on $\SSS(\R^2)$ by
$I_m\psi=\psi$ extends by density uniquely to
\begin{equation}
I_m:\: W^{1,2}(\R^2) \,\to\, L^2(m):= L^2(\R^2,\D m) \,;
\end{equation}
abusing notation we employ the same symbol for a continuous
function and the corresponding equivalence classes in both
$L^2(\R^2)$ and $L^2(m)$. The inequality (\ref{basiccond}) extends
to $W^{1,2}(\R^2)$ with $\psi$ replaced by $I_m\psi$ at the
left-hand side. This makes it possible to introduce the following
quadratic form,
\begin{equation} \label{Hamform}
\EE_{-\alpha m}(\psi,\phi) :=  \int_{\R^2}
\overline{\nabla\psi(x)} \nabla\phi(x) \, \D x -\alpha \int_{\R^2}
(I_m\bar\psi)(x) (I_m\phi)(x)\, \D m(x)\,,
\end{equation}
with the domain $W^{1,2}(\R^2)$; it is straightforward to see that
under the condition (\ref{basiccond}) it is closed and below
bounded, with $C_0^{\infty}(\R^2)$ as a core, and thus associated
with a unique self-adjoint operator.  Furthermore,
(\ref{basiccond}) is satisfied with any $a>0$ provided $m$ belongs
to the generalized Kato class,
\begin{equation} \label{Kato}
\lim_{\epsilon\to 0}\: \sup_{x\in\R^2}\, \int_{B(x,\epsilon)} |\ln
|x\!-\!y|| \, \D m(y) = 0\,,
\end{equation}
where $B(x,\epsilon)$ is the ball of radius $\epsilon$ and center
$x$. Choosing now for $m$ the Dirac measure supported by the curve
one can check easily that the condition (\ref{Kato}) is satisfied
under our assumptions about $\Gamma$, hence we may identify the
above mentioned self-adjoint operator with the formal one given by
(\ref{formal}).

The described definition applies naturally to a much wider class
of perturbations than we need here. Since $\Gamma$ is supposed to
be smooth, with the normal defined everywhere, we can define
$H_{\alpha,\Gamma}$ alternatively through boundary conditions.
Specifically, it acts as $-\Delta\psi$ on any $\psi$ from the
domain consisting of functions which belong to
$W^{2,2}(\mathbb{R}^2\setminus\Gamma)$, they are continuous at the
curve $\Gamma$ and their normal derivatives have a jump there,
 $$ 
 {\partial\psi(x)\over\partial n_+} -
 {\partial\psi(x)\over\partial n_-} =
  -\alpha \psi(x) \quad \mathrm{for}\quad x=\Gamma(s)\,,\; \forall
  s\in[0,L]\,.
 $$ 
It is straightforward to check that such an operator is e.s.a. and
its closure can be identified with (\ref{formal}) defined in the
above described way \cite{BEKS}. The advantage of the second
definition is that it has an illustrative meaning which
corresponds well to the concept of a $\delta$ interaction in the
cross cut of the curve.

Since the curve is finite, by \cite{BEKS, BT} we have
$\sigma_\mathrm{ess}(H_{\alpha,\Gamma})= [0,\infty)$ while the
discrete spectrum is nonempty and finite, so that
 $$ 
 \epsilon_1 \equiv \epsilon_1(\alpha,\Gamma):= \inf \sigma
 \left(-\Delta_{\alpha,\Gamma}\right)<0\,;
 $$ 
we ask for which $\Gamma$ the principal eigenvalue is maximal. The
main result of this paper is a partial answer to this question,
namely:

 \begin{theorem} \label{mainthm}
 Within the specified class of curves, $\epsilon_1(\alpha,\Gamma)$
 is for any fixed $\alpha>0$ and $L>0$ locally sharply maximized
 by a circle.
 \end{theorem}

\noindent While we do not give a general answer here, we suggest
what it should be.

 \begin{conjecture} \label{mainconj}
 The circle is a sharp global maximizer, even under weaker
 regularity assumptions.
 \end{conjecture}


\setcounter{equation}{0}
\section{Birman-Schwinger reformulation} \label{bs}

For operators associated with the quadratic form (\ref{Hamform})
one can establish a generalized Birman-Schwinger principle -- we
refer to \cite{BEKS} for a detailed discussion. In particular, if
$k^2$ belongs to the resolvent set of $H_{\alpha,\Gamma}$ we put
$R^k_{\alpha,\Gamma} := (H_{\alpha,\Gamma}-k^2)^{-1}$. The free
resolvent $R^k_0$ is defined for $\im k>0$ as an integral operator
in $L^2(\R^2)$ with the kernel
 $$ 
 G_k(x\!-\!y) = {i\over 4}\, H_0^{(1)} (k|x\!-\!y|)\,.
 $$ 
Next we have to introduce embedding operators associated with
$R^k_0$. Let $\mu, \nu$ be arbitrary positive Radon measures on
$\R^2$ with $\mu(x)= \nu(x) =0$ for any $x\in\R^2$. By
$R^k_{\nu,\mu}$ we denote the integral operator from
$L^2(\mu):=L^2(\R^2,\D\mu)$ to $L^2(\nu)$ with the kernel $G_k$,
in other words we suppose that
$$ R^k_{\nu,\mu} \phi = G_k \ast \phi\mu $$
holds $\nu$-a.e. for all $\phi\in D(R^k_{\nu,\mu}) \subset
L^2(\mu)$. In our case the two measures will be the Dirac measure
supported by $\Gamma$, denoted by $m$ if necessary,  and the
Lebesgue measure $\D x$ on $\R^2$, in different combinations. With
this notation one can express the generalized BS principle as
follows:
\begin{proposition} \label{bsprop}
(i) There is a $\kappa_0>0$ such that the operator $I-\alpha
R^{i\kappa}_{m,m}$ on $L^2(m)$ has a bounded inverse for any
$\kappa \ge \kappa_0$. \\ [1mm]
(ii) Let $\im k>0$. Suppose that $I-\alpha R^k_{m,m}$ is
invertible and the operator
$$ R^k := R_0^k + \alpha R^k_{\D x,m} [I-\alpha R^k_{m,m}]^{-1}
R^k_{m,\D x} $$
from $L^2(\R^2)$ to $L^2(\R^2)$ is everywhere defined. Then $k^2$
belongs to $\rho(H_{\alpha,\Gamma})$ and $(H_{\alpha,\Gamma}
-k^2)^{-1}= R^k$.
\\ [1mm]
(iii) $\:\dim\ker(H_{\alpha,\Gamma}-k^2) = \dim\ker(I-\alpha
R^k_{m,m})$ for any $k$ with $\im k>0$. \\ [1mm]
(iv) an eigenfunction of $H_{\alpha,\Gamma}$ associated with such
an eigenvalue $k^2$ can be written as
 $$ \psi(x)= \int_0^L R^k_{\D x,m}(x,s) \phi(s)\, \D s\,, $$
where $\phi$ is the corresponding eigenfunction of $\alpha
R^k_{m,m}$ with the eigenvalue one.
\end{proposition}

\noindent \textsl{Proof} of (i)-(iii) is given in \cite{BEKS}, for
(iv) see \cite{Po}. \QED \vspace{.5em}

Denoting conventionally $k=i\kappa$ with $\kappa>0$ as
corresponding to the bound-state energy $-\kappa^2$, we can thus
rephrase our problem as a search for solutions to the
integral-operator equation
 \begin{equation} \label{bsform}
 \RR_{\alpha,\Gamma}^\kappa \phi= \phi\,, \quad
 \RR_{\alpha,\Gamma}^\kappa(s,s'):= \frac{\alpha}{2\pi}
 K_0\big(\kappa|\Gamma(s) \!-\!\Gamma(s')|\big)\,,
 \end{equation}
on $L^2([0,L])$, where $K_0$ is Macdonald function. Referring
again to \cite{BEKS} and \cite{Po} we find that the
operator-valued function $\kappa\mapsto
\RR_{\alpha,\Gamma}^\kappa$ is strictly decreasing in $(0,\infty)$
and $\|\RR_{\alpha,\Gamma}^\kappa\|\to 0$ as $\kappa\to\infty$. In
fact the two properties can be checked also directly. The first
one follows from the one-to-one correspondence of the eigenvalue
branches (as functions of $\kappa$) to those of
$H_{\alpha,\Gamma}$ which are obviously strictly monotonous as
functions of $\alpha$; the second one in turn comes from the
explicit form of the kernel together with the dominated
convergence theorem.

Next we use the fact that the maximum eigenvalue of
$\RR_{\alpha,\Gamma}^\kappa$ is simple. This conclusion results
from the following considerations: the kernel of the operator is
by (\ref{bsform}) strictly positive, so
$\RR_{\alpha,\Gamma}^\kappa$ is positivity improving. It further
means that for any nonzero $\phi,\chi\ge 0$ the functions
$\RR_{\alpha,\Gamma}^\kappa\phi,\, \RR_{\alpha,
\Gamma}^\kappa\chi$ are also strictly positive. Hence $(\phi,
(\RR_{\alpha,\Gamma}^\kappa)^2\chi)\ne 0$, and as a consequence,
$\RR_{\alpha,\Gamma}^\kappa$ is ergodic; then the claim follows
from Thm.~XIII.43 of \cite{RS}. In view of
Proposition~\ref{bsprop}(iii) the ground state of
$H_{\alpha,\Gamma}$ is, of course, also simple.

If $\Gamma$ is a circle the operator $H_{\alpha,\CC}$ has a full
rotational symmetry, so the corresponding eigenspace supports a
one-dimensional representation of the group $O(2)$. Let us denote
the ground-state eigenfunction of $H_{\alpha,\CC}$ as
$-\tilde\kappa_1^2\,$ (we will use tilde to distinguish quantities
referring to the circle). The correspondence between the
eigenfunctions given by Proposition~\ref{bsprop}(iv) then requires
that the respective eigenfunction of
$\RR_{\alpha,\CC}^{\tilde\kappa_1}$ corresponding to the unit
eigenvalue is constant; we can choose it as $\tilde \phi_1(s)=
L^{-1/2}$. Then we have
 $$ 
 \max \sigma(\RR_{\alpha,\CC}^{\tilde\kappa_1}) = (\tilde\phi_1,
 \RR_{\alpha,\CC}^{\tilde\kappa_1} \tilde\phi_1) = \frac1L
 \int_0^L \int_0^L
 \RR_{\alpha,\CC}^{\tilde\kappa_1}(s,s') \,\D s\D s'\,,
 $$ 
and on the other hand, for the same quantity referring to a
general $\Gamma$ a simple variational estimate gives
 $$ 
 \max \sigma(\RR_{\alpha,\Gamma}^{\tilde\kappa_1}) \ge (\tilde\phi_1,
 \RR_{\alpha,\Gamma}^{\tilde\kappa_1} \tilde\phi_1) = \frac1L
 \int_0^L \int_0^L
 \RR_{\alpha,\Gamma}^{\tilde\kappa_1}(s,s') \,\D s\D s'\,.
 $$ 
Hence to check that the circle is a maximizer it sufficient to
show that
 \begin{equation} \label{Greenineq}
 \int_0^L \int_0^L
 K_0\big(\kappa|\Gamma(s) \!-\!\Gamma(s')|\big) \,\D s\D s' \ge
 \int_0^L \int_0^L
 K_0\big(\kappa|\CC(s) \!-\!\CC(s')|\big) \,\D s\D s'
 \end{equation}
holds \emph{for all} $\kappa>0$ and $\Gamma$ of the considered
class, or at least for $\Gamma$ in the vicinity of $\CC$ to prove
the local result in Theorem~\ref{mainthm}. Since the kernel is
symmetric w.r.t. the two variables, we can replace the double
integral by $2\int_0^L \D s\int_0^s\D s'$. By another simple
change of variables we find that the above claim is equivalent to
positivity of the functional
 $$ 
 F_\kappa(\Gamma):= \int_0^{L/2} \D u \int_0^L \D s \bigg[
 K_0\big(\kappa|\Gamma(s\!+\!u) -\Gamma(s)|\big) -
 K_0\big(\kappa|\CC(s\!+\!u) -\CC(s)|\big) \bigg]\,,
 $$ 
where the second term in the integrand is, of course, independent
of $s$ being equal to $K_0\big(\frac{\kappa L}{\pi} \sin \frac{\pi
u}{L}\big)$. Now we employ the (strict) convexity of $K_0$ which
yields by means of the Jensen inequality the following estimate,
 $$ 
 \frac 1L \,F_\kappa(\Gamma)\ge \int_0^{L/2} \left[
 K_0\left( \frac{\kappa}{L} \int_0^L |\Gamma(s\!+\!u) -\Gamma(s)| \D s
 \right) - K_0\left(\frac{\kappa L}{\pi} \sin \frac{\pi
 u}{L}\right) \right]\, \D u\,,
 $$ 
where the inequality is sharp unless $\int_0^L |\Gamma(s\!+\!u)
-\Gamma(s)| \D s$ is independent of~$s$. Finally, we observe that
$K_0$ is decreasing in $(0,\infty)$, hence it is sufficient to
check the inequality
 \begin{equation} \label{suffic}
 \int_0^L |\Gamma(s\!+\!u) -\Gamma(s)|\, \D s \,\le\,
 \frac{L^2}{\pi} \sin \frac{\pi u}{L}
 \end{equation}
for all $u\in(0,\frac12 L]$ and to show that is sharp unless
$\Gamma$ is a circle.


\setcounter{equation}{0}
\section{Mean-chord inequalities} \label{geom}

The inequality (\ref{suffic}) to which we have reduced our problem
can be regarded as an element of a wider family which we are now
going to describe. Let $\Gamma:\: [0,L]\to \R^2$ be again a loop
in the plane; for the moment we do not specify its regularity
properties. Let us consider all the arcs of $\Gamma$ having length
$u\in (0,\frac12 L]$. The mentioned inequalities are the following
 \begin{eqnarray}
 C_L^p(u):&\quad\;\; \int_0^L |\Gamma(s\!+\!u) -\Gamma(s)|^p\,
 \D s \,\le\, \frac{L^{1+p}}{\pi^p} \sin^p \frac{\pi u}{L}\,,&
 \; p>0\,, \label{C+p} \\
 C_L^{-p}(u):& \int_0^L |\Gamma(s\!+\!u) -\Gamma(s)|^{-p}\,
 \D s \,\le\, \frac{\pi^p L^{1-p}}{\sin^p \frac{\pi u}{L}}\,,&
 \; p>0\,. \label{C-p}
 \end{eqnarray}
They have also a discrete counterpart for an equilateral polygon
$\PP_N$ of $N$ vertices and side length $\ell>0$. Let $\{y_n\}$ be
the family of its vertices, where the index values are identified
modulo $N$; then we introduce
 \begin{eqnarray}
 D_{N,\ell}^p(m):&\; \sum_{n=1}^N |y_{n+m}-y_n|^p\,
 \le\, \frac{N\ell^p \sin^p \frac{\pi m}{N}}{\sin^p \frac{\pi}{N}} \,,&
 \; p>0\,, \label{D+p} \\
 D_{N,\ell}^{-p}(m):& \sum_{n=1}^N |y_{n+m}-y_n|^{-p}\,
 \le\, \frac{N \sin^p \frac{\pi }{N}}{\ell^p \sin^p \frac{\pi m}{N}}\,,&
 \; p>0\,, \label{D-p}
 \end{eqnarray}
for any $m=1,\dots, [\frac12 N]$, where $[\cdot]$ denotes as usual
the entire part.

In all the cases the right-hand side corresponds, of course, to
the case with maximal symmetry, i.e. to the circle and regular
polygon $\tilde\PP_N$, respectively. \emph{We conjecture} that
without regularity restrictions $C_L^{\pm p}(u)$ \emph{holds for
any $p\le 2$ and the same is true for} $D_{N,\ell}^{\pm p}(m)$,
and furthermore, we expect the inequalities \emph{to be sharp
unless $\Gamma=\CC$ or $\PP_N=\tilde\PP_N$, respectively}. In the
polygon case it is clear that the claim may not be true for $p>2$
as the example of a rhomboid shows: $D_{4,\ell}^p(2)$ is
equivalent to $\sin^p\phi +\cos^p\phi \ge 1$ for $0<\phi<\pi$. We
are unable at this moment to demonstrate the inequalities
(\ref{C+p})--(\ref{D-p}) in full generality; below we will present
a few particular cases.

It is obvious that the inequalities have a scaling property, so
without loss of generality one can assume, e.g., $L=1$ and
$\ell=1$; in such a case we drop the corresponding symbol from the
label. If necessary we can include also the case $p=0$ when the
inequalities turn into trivial identities.

\begin{proposition} \label{monotone}
$C_L^p(u)\Rightarrow C_L^{p'}(u)$ and $D_{N,\ell}^p(m)\Rightarrow
D_{N,\ell}^{p'}(m)$ if $p>p'>0$.
\end{proposition}
{\sl Proof:} The claim follows from the convexity of $x\mapsto
x^\alpha$ in $(0,\infty)$ for $\alpha>1$,
 \begin{eqnarray*}
 \frac{L^{1+p}}{\pi^p} \sin^p \frac{\pi u}{L} &\!\ge\!&
 \int_0^L \left(|\Gamma(s\!+\!u) -\Gamma(s)|^{p'} \right)^{p/p'}\,
 \D s \\ &\!\ge\!&
 L \left( \frac1L \int_0^L |\Gamma(s\!+\!u) -\Gamma(s)|^{p'}\,
 \D s \right)^{p/p'}\,.
 \end{eqnarray*}
It is now sufficient to take both sides to the power $p'/p$; in
the same way one checks the second implication. \QED \vspace{.5em}

\begin{proposition} \label{inverse}
$C_L^p(u)\Rightarrow C_L^{-p}(u)$ and $D_{N,\ell}^p(m)\Rightarrow
D_{N,\ell}^{-p}(m)$ for any $p>0$.
\end{proposition}
{\sl Proof:} The Schwarz inequality implies
 $$ 
 \int_0^L |\Gamma(s\!+\!u) -\Gamma(s)|^{-p}\, \D s \ge
 \frac{L^2}{\int_0^L |\Gamma(s\!+\!u) -\Gamma(s)|^p\, \D s} \ge
 \frac{L^2\pi^p}{L^{1+p}\sin^p \frac{\pi u}{L}}\,,
 $$ 
and similarly for the polygon case. \QED \vspace{.5em}

These simple relations mean that to check the above stated
conjecture one needs only to verify $C^2(u)$ and $D_N^2(m)$. We
will address the continuous case in the next section, here we
notice that the results of \cite{Ex} in combination with the last
two propositions leads to the following conclusions:

 \begin{theorem} \label{polygon1}
 (a) $\;D^1_{N,\ell}(m)$ holds locally for any $N$ and $m=1,\dots,
 [\frac12 N]$, i.e. in a vicinity of the regular polygon, and
 consequently, $D^{\pm p}_{N,\ell}(m)$ holds locally for any $p\in
 (0,1]$. \\ [.2em]
 (b) $\;D^1_{N,\ell}(2)$ holds globally for any $N$, and so does
 $D^{\pm p}_{N,\ell}(2)$ for each $p\in (0,1]$.
 \end{theorem}


\setcounter{equation}{0}
\section{Proof of Theorem~\ref{mainthm}} \label{proof}

After this interlude let us return to our main problem. Notice
first that our regularity hypothesis allows us to characterize
$\Gamma$ by its (signed) curvature $\gamma:= \dot\Gamma_2
\ddot\Gamma_1- \dot\Gamma_1\ddot\Gamma_2$ which is by assumption a
piecewise continuous function in $[0,L]$. The advantage is that
$\gamma$ specifies uniquely the equivalence class related by
Euclidean transformations which can be represented by
 \begin{equation} \label{param}
 \Gamma(s)= \left( \int_0^s \cos\beta(s')\, \D s',
 \int_0^s \sin\beta(s')\, \D s' \right)\,,
 \end{equation}
where $\beta(s):= \int_0^s \gamma(s')\, \D s'$ is the bending
angle relative to the tangent at the chosen initial point, $s=0$.
To ensure that the curve is closed, the conditions
 \begin{equation} \label{close}
 \int_0^L \cos\beta(s')\, \D s' =
 \int_0^L \sin\beta(s')\, \D s' = 0
 \end{equation}
must be satisfied. Using this parametrization we can rewrite the
left-hand side of the inequality (\ref{C+p}) in the form
 $$ 
 \int_0^L \left[ \left( \int_s^{s+u} \cos\beta(s')\,
 \D s'\right)^2 + \left( \int_s^{s+u} \sin\beta(s')\,
 \D s'\right)^2 \right]^{p/2} \!\D s:=c^p_\Gamma(u)\,,
 $$ 
or equivalently
 $$ 
 c^p_\Gamma(u) = \int_0^L \D s \left[ \int_s^{s+u} \D s'
 \int_s^{s+u} \D s'' \cos(\beta(s')-\beta(s'') \right]^{p/2}.
 $$ 
By Proposition~\ref{monotone} it is sufficient to check that the
quantity $c^2_\Gamma(u)$ is maximized by the circle, i.e. by
$\beta(s)= \frac{2\pi s}{L}$. Rearranging the integrals we get
 \begin{eqnarray*}
 && c^2_\Gamma(u) =
 \int_0^L \D s' \int_{s'-u}^{s'+u} \D s''
 \int_{\max{\{s'-u,s''-u\}}}^{\min{\{s',s''\}}} \D s\,
 \cos(\beta(s')-\beta(s'')) \\ && =
 \int_0^L \D s' \int_{s'\!-u}^{s'\!+u} \!\D s''\,
 [\min{\{s',s''\}}-\max{\{s'\!-u,s''\!-u\}}]
 \cos(\beta(s')\!-\!\beta(s'')),
 \end{eqnarray*}
or
 $$ 
 c^2_\Gamma(u)= \int_0^L \D s' \int_{s'-u}^{s'+u} \D s''\,
 \left[ u-|s'\!-s''|\right]\,\cos(\beta(s')-\beta(s''))\,.
 $$ 
Next we change the integration variables to $x:=s'\!-s''$ and $z:=
\frac12(s'\!+s'')$,
 $$ 
 c^2_\Gamma(u)= \int_{-u}^u \D x\, (u-|x|) \int_0^L \D z\,
 \cos\left(\beta(z+\frac12 x)-\beta(z-\frac12 x)\right)\,,
 $$ 
and since the functions involved are even w.r.t. $x$ we finally
get
 \begin{equation} \label{c^2}
 c^2_\Gamma(u)= 2\int_0^u \D x\, (u-x) \int_0^L \D z\,
 \cos\left(\int_{z-\frac12 x}^{z+\frac12 x} \gamma(s)\,
 \D s \right)\,.
 \end{equation}
As a certain analogy to Theorem~\ref{polygon1}(b) we can prove the
sought global inequality in case when the curve arcs in question
are sufficiently short and/or the tangent vector direction does
change too fast.

\begin{proposition} \label{part global}
Suppose that $\Gamma$ has no self-intersections and the inequality
$\beta(z+\frac12 u)-\beta(z-\frac12 u) \le \frac12 \pi$ is valid
for all $z\in[0,L]$, then $C_L^2(u)$ holds.
\end{proposition}
{\sl Proof:} We employ concavity of cosine in $(0, \frac12\pi)$
obtaining
 \begin{eqnarray*}
 c^2_\Gamma(u) &\!\le\!& 2L\int_0^u \D x\, (u-x)
 \cos\left(\frac1L \int_0^L \D z\,
 \int_{z-\frac12 x}^{z+\frac12 x} \gamma(s)\, \D s \right) \\
 &\!=\!& 2L\int_0^u \D x\, (u-x)
 \cos\left(\frac1L \int_0^L \D s\,\gamma(s)
 \int_{s-\frac12 x}^{s+\frac12 x} \D z \right) \\
 &\!=\!& 2L\int_0^u \D x\, (u-x) \cos \frac{2\pi x}{L} =
 \frac{L^3}{\pi^2}\, \sin^2 \frac{\pi u}{L}\,,
 \end{eqnarray*}
since $\int_0^L \gamma(s)\,\D s = \pm 2\pi$ for a curve without
self-intersections. Moreover, the function $z\mapsto
\int_{z-\frac12 x}^{z+\frac12 x} \gamma(s)\, \D s$ is constant for
$x\in (0,u)\,$ \emph{iff} $\,\gamma(\cdot)$ is constant, hence the
circle corresponds to a sharp maximum. \QED \vspace{.5em}

This result, however, does not help us with our main problem,
because we need the inequality to be valid for all arc lengths. As
indicated before, we can prove a local result which will imply
Theorem~\ref{mainthm}.

 \begin{theorem} \label{smalldef}
 Under the regularity assumptions of Sec.~\ref{bs}, the inequality
 $C^2_L(u)$ holds locally for any $L>0$ and $u\in (0,\frac12 L]$,
 and consequently, $C^{\pm p}_L(u)$ holds locally for any $p\in
 (0,2]$.
 \end{theorem}
{\sl Proof:} Gentle deformations of a circle can be characterized
by the curvature
 $$ 
 \gamma(s)= \frac{2\pi}{L} +g(s)\,,
 $$ 
where $g$ is a piecewise continuous functions which is small in
the sense that $\|g\|_\infty\ll L^{-1}$ and satisfies the
condition $\int_0^L g(s)\, \D s=0$. The function in the last
integral of (\ref{c^2}) can be then expanded as
 $$ 
 \cos\frac{2\pi x}{L} - \sin\frac{2\pi x}{L}
 \int_{z-\frac12 x}^{z+\frac12 x} g(s)\, \D s -
 \frac12 \cos\frac{2\pi x}{L}
 \left( \int_{z-\frac12 x}^{z+\frac12 x} g(s)\, \D s \right)^2
 + \OO(g^3)\,,
 $$ 
where the error term is a shorthand for $\OO(\|Lg\|_\infty^3)$.
Substituting this expansion into (\ref{c^2}) we find that the term
linear in $g$ vanishes, because
 $$ 
 \int_0^L \D z \int_{z-\frac12 x}^{z+\frac12 x} g(s)\, \D s =
 \int_0^L \D s\, g(s) \int_{s-\frac12 x}^{s+\frac12 x} \D z =
 0\,,
 $$ 
and thus
 \begin{equation} \label{c^2exp}
 c^2_\Gamma(u)= \frac{L^3}{\pi^2} \sin^2 \frac{\pi u}{L}
 - I_g(u) + \OO(g^3)\,,
 \end{equation}
where
 $$ 
 I_g(u):= \int_0^u \D x\, (u-x)\, \cos\frac{2\pi x}{L}
 \int_0^L \D z\, \left(\int_{z-\frac12 x}^{z+\frac12 x} g(s)\,
 \D s \right)^2.
 $$ 
We need to show that $I_g(u)>0$ unless $g=0$ identically. Notice
that for $u\le \frac14 L$ this property holds trivially. For $u\in
(\frac14 L, \frac12 L]$ we use the fact that $g$ is periodic and
piecewise continuous, so we can write it through its Fourier
series
 $$ 
 g(s) = \sum_{n=1}^\infty \left( a_n \sin\frac{2\pi ns}{L}
 + b_n \cos\frac{2\pi ns}{L} \right)
 $$ 
with the zero term missing, where $\sum_n(a_n^2+b_n^2)$ is finite
(and small). Using
 $$ 
 \int_{z-\frac12 x}^{z+\frac12 x} g(s)\, \D s =
 \frac{L}{\pi} \sum_{n=1}^\infty \frac1n\,
 \left( a_n \sin\frac{2\pi nz}{L}
 + b_n \cos\frac{2\pi nz}{L} \right) \sin\frac{\pi nx}{L}
 $$ 
together with the orthogonality of the Fourier basis we find
 $$ 
 I_g(u)= \int_0^{u} \D x\, (u-x)\, \cos\frac{2\pi x}{L}
 \sum_{n=1}^\infty \frac{L^3}{2\pi^2}\, \frac{a_n^2+b_n^2}{n^2}\,
 \sin\frac{\pi nx}{L}\,.
 $$ 
Since the summation and integration can be obviously interchanged,
we have
 \begin{equation} \label{I_gexpand}
 I_g(u)= \frac{L^5}{2\pi^4}\,
 \sum_{n=1}^\infty \, \frac{a_n^2+b_n^2}{n^2}\,
 F_n\left( \frac{\pi u}{L}\right)\,,
 \end{equation}
where
 $$ 
 F_n(v):= \int_0^{v} (v-y)\, \cos 2y\, \sin ny\: \D y\,.
 $$ 
These integrals are equal to
 \begin{eqnarray*}
 F_1(v) &\!=\!& \frac{1}{18} (\,9\sin v -\sin 3v -6v)\,, \\
 F_2(v) &\!=\!& \frac{1}{32} (\,4v -\sin 4v) \,, \\
 F_n(v) &\!=\!& \frac{nv}{n^2-4} - \frac{\sin(n-2)v}{2(n-2)^2}
 - \frac{\sin(n+2)v}{2(n+2)^2} \,, \qquad n\ge 3\,.
 \end{eqnarray*}
Using the fact that $\sin x<x$ for $x>0$ we see immediately that
$F_n(v)>0$ for $v>0$ and $n\ge 2$. On the other hand, $F_1(v)$ has
in the interval $(0,\frac{\pi}{2})$ a single positive maximum, at
some $v> \frac{\pi}{4}$, from which it decreases to the value
$F_1(\frac{\pi}{2})= \frac{1}{18}(10-3\pi)>0$. Summing up this
argument, we have found that the quantity (\ref{I_gexpand}) is
positive unless all the coefficients $a_n,b_n$ are zero. \QED

\begin{remark} \label{closecurve}
{\rm One may wonder what happened with the closedness requirement
(\ref{close}). As the argument shows we were able to demonstrate
the claim using only the weaker property that $\beta(0)=\beta(L)$.
This is possible, of course, for small deformations only! As an
illustration, consider $\Gamma$ in the form of an ``overgrown
paperclip'' which satisfies the condition $\beta(0)=\beta(L)$ but
not (\ref{close}), i.e. a line segment with two U-turns at the
ends. Making the latter short one can get $c^2_\Gamma(\frac12 L)$
arbitrarily close to $\frac13 L^3$ which is larger than
$L^3/\pi^2$. }
\end{remark}


\setcounter{equation}{0}
\section{Extensions and conclusions} \label{concl}

To support our expectations that the result given in
Theorem~\ref{mainthm} holds globally and under weaker regularity
assumptions, consider a simple example.

\begin{example} \label{lens+appla}
{\rm Let $\Gamma$ be a curve consisting of two circular segments
of radius $R> \frac{L}{4\pi}$, i.e. it is given by the equations
 \begin{equation} \label{lens}
 \left( x\pm R\cos\frac{L}{2R} \right)^2 + y^2 = R^2 \qquad
 \mathrm{for}\quad \pm x\ge 0\,.
 \end{equation}
For $R> \frac{L}{2\pi}$ it is ``lens-shaped'', for $\frac{L}{4\pi}
<R< \frac{L}{2\pi}$ ``apple-shaped''; it is not smooth except in
the trivial case of a circle, $R= \frac{L}{2\pi}$. The curvature
of this $\Gamma$ equals
 $$ 
 \gamma(s)= \frac1R + \left(\pi- \frac{L}{2R}\right) (\delta(s)
 +\delta(s-L/2))\,,
 $$ 
hence
 $$ 
 c^2_\Gamma(u)= 2\int_0^u \D x\, (u-x) \left[ (L-2x)
 \cos\frac{x}{R} - 2x \cos \frac{L-2x}{2R} \right] \D x\,,
 $$ 
and evaluating the integral, we arrive at
 $$ 
 c^2_\Gamma(u)= 8R^3 \left\{ \frac{L}{2R} \sin^2 \frac{u}{2R} +
 4\left( \frac{u}{2R} \cos \frac{u}{2R} - \sin \frac{u}{2R} \right)
 \cos \frac{L}{4R}\, \cos \frac{L-2u}{4R} \right\}\,.
 $$ 
This function has for each $u\in(0,\frac12 L]$ a maximum at $R=
\frac{L}{2\pi}$ and one can check directly that its value is
smaller for any other $R$. In particular, in the limit $R\to
\infty$ we have $c^2_\Gamma(u)\to Lu^2-\frac43 u^3$ as one can
find also directly with the ``lens'' degenerate into a double line
segment; this value is less than $\frac{L^3}{\pi^2} \sin^2
\frac{\pi u}{L}$ because $\sin^2x > x^2- \frac{4}{3\pi} x^3$ holds
in $(0,\frac{\pi}{2})$. }
\end{example}

To summarize our discussion, to prove Conjecture~\ref{mainconj} it
is sufficient to verify the inequality $C_L^p(u)$ for some $p\ge
1$ under appropriate regularity hypothesis. Naturally, one can ask
also about ground-state maximizer in smaller families of curves
$\Gamma$ which do not contain the circle; examples could be
polygonal loops with a fixed or limited number of vertices, or
various prescribed compositions of arcs belonging to specific
classes, circular, elliptic, parabolic, etc. Obviously a
reasonable strategy is to look first for curves as close to the
circle as possible within the given class. Sometimes one expects
that the answer will be the curve with maximum symmetry as in the
polygon case, in other situations it may not be true.

Another, and maybe more important extension of the present problem
concerns a maximizer for the generalized Schr\"odinger operator in
$\R^3$ with an attractive $\delta$ interaction supported by a
closed surface of a fixed area $A$, and its generalization to
closed hypersurfaces of codimension one in $\R^d,\, d>3$. In the
case of $d=3$ we have a heuristic argument relying on \cite{Ex2,
EHL} similar to that used in the introduction which suggests that
the problem is solved by the sphere provided the discrete spectrum
is not empty, of course, which is a nontrivial assumption in this
case -- for properties of the corresponding operators see
\cite{AGS}. The Birman-Schwinger reduction of the problem similar
to that of Sec.~\ref{bs} can be performed again and the task is
thus reduced to verification of a geometric inequality analogous
to (\ref{C+p}) which we can label as $C_A^{d,p}(u)$. We will
discuss this problem in a following paper.


\subsection*{Acknowledgments}

The research has been partially supported
by the ASCR Grant Agency within the project A100480501.

\end{document}